\documentclass{webofc}
\usepackage[varg]{txfonts}   
\usepackage{xcolor}
\usepackage[colorlinks=True, citecolor=blue, urlcolor=blue, linkcolor=blue]{hyperref}
\begin{document}
\title{Quarkonium transport in weakly and strongly coupled plasmas}

\author{\firstname{Govert} \lastname{Nijs}\inst{1,2}\fnsep\thanks{\email{govert.nijs@cern.ch}} \and
        \firstname{Bruno} \lastname{Scheihing-Hitschfeld}\inst{2}\fnsep\thanks{\email{bscheihi@mit.edu}, speaker at QM 2023} \and
        \firstname{Xiaojun} \lastname{Yao}\inst{3}\fnsep\thanks{\email{xjyao@uw.edu}}
}

\institute{Theoretical Physics Department, CERN, CH-1211 Gen\`eve 23, Switzerland 
\and
           Center for Theoretical Physics, Massachusetts Institute of Technology, Cambridge, MA  02139, USA 
\and
           InQubator for Quantum Simulation, Department of Physics, University of Washington, Seattle, WA 98195, USA
          }

\fnsep{\!\!\!\!\!MIT-CTP/5661,~IQuS@UW-21-067, CERN-TH-2023-245}

\abstract{
We report on progress in the nonperturbative understanding of quarkonium dynamics inside a thermal plasma. The time evolution of small-size quarkonium is governed by two-point correlation functions of chromoelectric fields dressed with an adjoint Wilson line, known in this context as generalized gluon distributions (GGDs). 
The GGDs have been calculated in both weakly and strongly coupled plasmas by using perturbative and holographic methods. Strikingly, the results of our calculations for a strongly coupled plasma indicate that the quarkonium dissociation and recombination rates vanish in the transport descriptions that assume quarkonium undergoes Markovian dynamics. However, this does not imply that the dynamics is trivial.
As a starting point to explore the phenomenological consequences of the result at strong coupling, we show a calculation of the $\Upsilon(1S)$ formation probability in time-dependent perturbation theory. This is a first step towards the development of a transport formalism that includes non-Markovian effects, which, depending on how close the as of yet undetermined nonperturbative QCD result of the GGDs is to the strongly coupled $\mathcal{N}=4$ SYM result, could very well dominate over the Markovian ones in quark-gluon plasma produced at RHIC and the LHC.
}
\maketitle
%
Quarkonium suppression in heavy ion collisions is an important probe of the quark-gluon plasma (QGP). Interpreting the full wealth of data obtained from the experiments and the extraction of QGP properties require a systematic theoretical understanding of the evolution of heavy quarks and quarkonia in the nearly thermal and strongly coupled plasma.

Recent studies that combine the open quantum system framework and effective field theory techniques obtained such a systematic field-theoretical understanding for heavy quark-antiquark pairs ($Q\bar{Q}$) of small separation. More specifically, when the heavy quark mass $M$ is large (velocity $v$ is small) and the $Q\bar{Q}$ distance $r$ is much smaller than the relevant medium scales such as the inverse of temperature $1/T$, the time evolution of the $Q\bar{Q}$ can be described by potential nonrelativistic QCD (pNRQCD)~\cite{Brambilla:1999xf}, which is constructed by systematic expansions in $1/M$ and $r$. The relevant degrees of freedom are $Q\bar{Q}$ pairs that can be in color singlet or octet state. The singlet and octet states can transition to each other via a color dipole interaction. In addition to that, octet states carry an adjoint color charge and thus they develop an SU$(N_c)$ phase in their time evolution, which is described by an adjoint Wilson line. At the leading nontrivial order in the aforementioned expansions, quarkonium states are color singlet states. As such, quarkonium dissociation and recombination processes are described as singlet-octet transitions.

Several studies have shown~\cite{Binder:2021otw,Scheihing-Hitschfeld:2022xqx,Scheihing-Hitschfeld:2023tuz} that quarkonium dynamics in a thermal medium, both for weakly and strongly coupled mediums, is governed by the following correlators
\begin{align} 
\label{eq:g++-definition}
[g_{\rm adj}^{++}]^>(t) &\equiv \frac{g^2 T_F }{3 N_c}  \big\langle E_i^a(t)W^{ac}(t,+\infty) 
W^{cb}(+\infty,0) E_i^b(0) \big\rangle_T \, , \\ \label{eq:g---definition}
[g_{\rm adj}^{--}]^>(t) &\equiv \frac{g^2 T_F }{3 N_c} \big\langle W^{dc}(-i\beta - \infty, -\infty)
W^{cb}(-\infty,t)  E_i^b(t)
E_i^a(0)W^{ad}(0,-\infty)  \big\rangle_T  \, ,
\end{align}
which we refer to as \textit{Generalized Gluon Distributions} (GGDs) because they represent the effective distribution of gluon-like scatterers from the medium that can mediate a dipole transition between singlet and octet states. 

\section{The strongly coupled limit in $\mathcal{N}=4$ Yang-Mills theory}
\label{sec:strong-coupling}

The generalized gluon distributions in Eqs.~\eqref{eq:g++-definition} and~\eqref{eq:g---definition} are part of a family of mutually related correlators that take part in quarkonium physics, which can be found in~\cite{Scheihing-Hitschfeld:2023tuz}. Because these correlators are formulated purely in terms of gauge fields, it is possible to evaluate them in theories other than QCD, and doing so can shed light on the qualitative aspects that might be expected once a lattice QCD determination of them becomes available.

As a paradigmatic model for a strongly coupled plasma, we consider supersymmetric $\mathcal{N}=4$ Yang-Mills theory at finite temperature, in the large $N_c$ and strong coupling limit $\lambda = g^2 N_c \gg 1$, where calculations become feasible by means of the AdS/CFT correspondence. In this limit of the theory, the correlator of the family that is most simply formulated is the time-ordered correlator. We calculated it in~\cite{Nijs:2023dks}. It is given by
\begin{align}
     [g_{\rm adj}^{{\mathcal{T}}}](\omega) = \frac{g^2T_F}{3 N_c} \!\! \int_{-\infty}^\infty \!\!\! dt \, e^{i\omega t}  \langle \hat{\mathcal{T}} E_i^a(t) \mathcal{W}^{ab}_{[t,0]} E_i^b(0) \rangle_T =  \frac{ (\pi T)^3 \sqrt{\lambda} \, T_F}{{ 4}\pi} \left( \frac{-i}{F^-_{|\omega|}(0)} \frac{\partial^3 F^-_{|\omega|}}{\partial \xi^3}(0) \right) \, ,
\end{align}
where $F^-_\omega$ is the regular solution of the differential equation
\begin{align} \label{eq:F-thermal}
    \frac{\partial^2 F^-_\omega}{\partial \xi^2} - 2 \left[ \frac{1 + \xi^4}{\xi(1-\xi^4)} - \frac{i \Omega \xi^3}{1-\xi^4} \right] \frac{\partial F^-_\omega}{\partial \xi} + \left[ \frac{i \Omega \xi^2}{1-\xi^4} + \frac{\Omega^2 (1 - \xi^6) }{(1-\xi^4)^2} \right] F^-_\omega = 0 \, ,
\end{align}
with $\Omega = \omega/(\pi T)$. When the medium is flowing relative to the heavy quark pair, 
the effective temperature experienced by quarkonium gets modified by a multiplicative factor given by the square root of the Lorentz boost factor $T \to T_{\rm eff} = \sqrt{\gamma} \, T$~\cite{Nijs:2023dbc}.


However, as discussed in~\cite{Nijs:2023dks,Nijs:2023ucw,Nijs:2023dbc}, the analytic structure of the time-ordered correlator defined by $F^-_{|\omega|}$ implies that both the zero frequency and the negative frequency parts of $[g_{\rm adj}^{++}](\omega)$ vanish. Therefore, the medium-induced dissociation and regeneration probabilities in recently developed transport formalisms for quarkonia, such as the quantum Brownian motion limit~\cite{Brambilla:2016wgg-2017zei} or the quantum optical limit~\cite{Yao:2018nmy}, vanish in such a plasma. We find the fact that both of these descriptions assume that the dynamics of quarkonium is Markovian is intimately related to this consequence, an assumption we will revisit in what follows.


\section{Beyond the Markovian approximation}
\label{sec:non-Markov}

To understand the dynamics of quarkonium in a strongly coupled plasma, it is therefore necessary to revisit the derivation of the transport formalisms where the generalized gluon distributions appear. To do this, we recall that the full quantum dynamics of the system, without any approximations, determines the (reduced) $Q\bar{Q}$ density matrix by evolving the full density matrix of the system and then tracing out the environment degrees of freedom (the QGP), i.e.,
\begin{align}
    \rho_{Q\bar{Q}}(t) = {\rm Tr}_{\rm QGP} \left[ U(t) \rho_{\rm tot}(t=0) U^\dagger(t) \right] \, , \label{eq:oqs-general}
\end{align}
where $\rho_{\rm tot}(t=0)$ is the initial density matrix of the whole system.

The small parameter $T/(Mv) \sim r T$ allows one to expand the time evolution operators $U(t)$ in Eq.~\eqref{eq:oqs-general} in this power counting parameter and keep only terms at the first nontrivial order. From this point forward, derivations of evolution equations for the reduced density matrix $\rho_{Q\bar{Q}}$ usually rely on additional assumptions beyond those required to set up the pNRQCD description of the heavy quark pair. For instance, in the quantum Brownian Motion limit~\cite{Brambilla:2016wgg-2017zei} it is assumed that $T \gg Mv^2 \sim \Delta E$, so that the typical time scale of the medium is much shorter than the characteristic time associated to the energy level splittings $\Delta E$ of quarkonium. This scale separation means that the dynamics will be determined by the zero frequency limit of $[g_{\rm adj}^{++}]^>(\omega)$, which vanishes in the strongly coupled $\mathcal{N}=4$ plasma. 
Another example is the quantum optical limit~\cite{Yao:2018nmy}, which is a semi-classical description applicable when $T \sim \Delta E$, where the off-diagonal elements of the density matrix are {grouped into the Wigner function}. As a consequence of the hierarchy of energy scales and the semi-classical limit, the contributions to in-medium quarkonium dynamics come only from the negative frequency part of $[g_{\rm adj}^{++}]^>(\omega)$, which also vanishes in the strongly coupled $\mathcal{N}=4$ plasma.

However, without these extra assumptions, a direct perturbative expansion of~\eqref{eq:oqs-general} gives a nonvanishing result. For definiteness, we consider an initial $Q\bar{Q}$ state in the color octet. One can then show that the probability that the $Q\bar{Q}$ pair is in a singlet state with quantum numbers $n,l$, after being in contact with a thermal bath between times $\tau_i$ and $\tau_f$, is given by
\begin{align}
    &\langle nl | \, \rho_{Q \bar{Q}}(\tau_f) \, | nl \rangle \\ &= \int_{\tau_i}^{\tau_f} \!\!\! d\tau_1 \int_{\tau_i}^{\tau_f} \!\!\! d\tau_2 \, [g_{\rm adj}^{--}]^{>}({\tau_2,\tau_1}) \, \langle nl | U^{\rm singlet}_{[\tau_f,\tau_1]} r_i U^{\rm octet}_{[\tau_1, \tau_i]} | \psi_0 \rangle \big( \langle nl | U^{\rm singlet}_{[\tau_f,\tau_2]} r_i U^{\rm octet}_{[\tau_2, \tau_i]} | \psi_0 \rangle \big)^\dagger \, , \nonumber
\end{align}
where $| \psi_0 \rangle$ is the initial wavefunction for the relative position coordinate of the $Q\bar{Q}$ pair in the octet state, and $U^{\rm singlet}_{[t,t']}$, $U^{\rm octet}_{[t,t']}$ are the one-body time evolution operators from time $t'$ to time $t$ for states in the singlet and octet representations, respectively, acting only on the wavefunction for the relative position coordinate between the heavy quark pair in each case. 

\begin{figure}[t]
    \centering
    \includegraphics[width=0.7\textwidth]{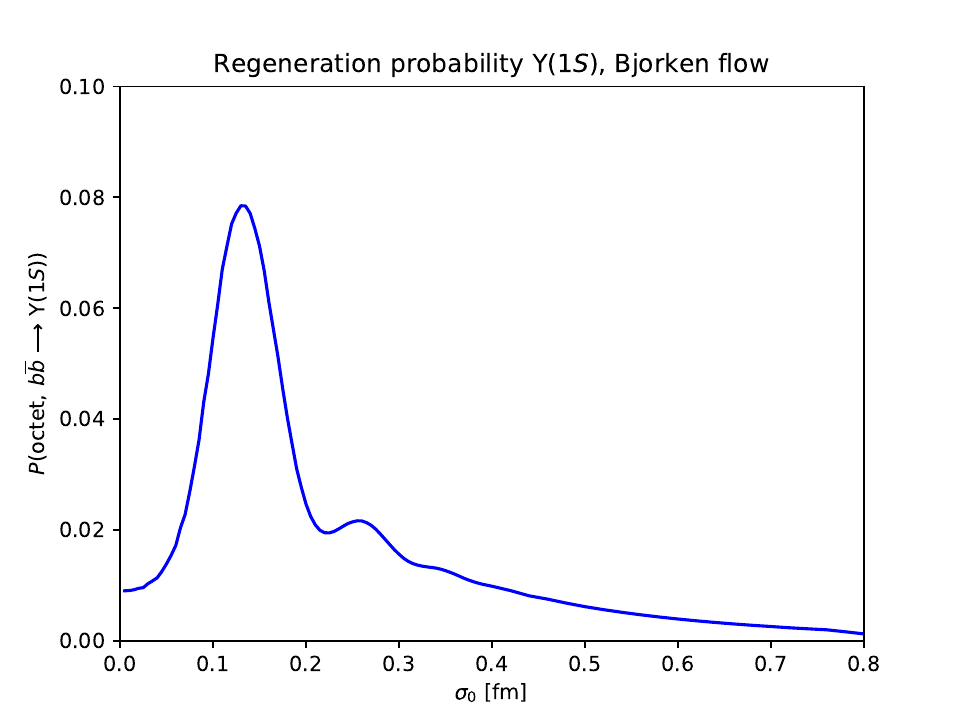}
    \caption{Regeneration/formation probability for an $\Upsilon(1S)$ state as a function of the initial separation $\sigma_0$ between the two heavy quarks. The temperature profile experienced by the heavy quark pair was set to be given by Bjorken flow scaling, $T(\tau) = (\tau_f/\tau)^{1/3} T_f $, with $T_f = 155 \, {\rm MeV}$, $\tau_i = 0.6 \, {\rm fm/c}$, and $\tau_f = 10 \, {\rm fm/c}$. The initial condition for the wavefunction in the radial component of the relative coordinate was given by $\psi_0(r) \propto r Y_{1m}(\theta, \varphi) \exp(- r^2/(2\sigma_0^2)) $, where $Y_{1m}$ is a spherical harmonic. The reason to choose $\ell=1$ as the initial state is that the transition to the $1S$ state is allowed by the dipole interaction of pNRQCD at the order we work in the EFT. The wavefunction is appropriately normalized to have unit probability, and the final result is averaged over $m$. }
    \label{fig:prob}
\end{figure}

To illustrate this formula, we plot it as a function of different initial conditions in Figure~\ref{fig:prob}. To do this, we need to specify an interaction potential model to construct the time evolution operators, which we take to be a Karsch-Mehr-Satz potential~\cite{kms:1998} for the singlet, and no potential for the octet. 
After specifying the model and the temperature of the system as a function of time, the only remaining parameter that needs to be specified in this description is the coupling constant, which for present purposes we set to $\lambda = 13$, corresponding to $g \approx 2.1$.

The result in Figure~\ref{fig:prob} shows many desirable properties. Most importantly, the formation probability is peaked around initial characteristic separations on the order of the size of the $\Upsilon(1S)$ state. The rest of the peak structure contains information about the in-medium excited states. Crucially, all of these features are within the applicability range of the EFT, $r T < 1$.

We stress that if either of the previously described transport formalisms had been used to calculate this probability with this correlator, the result would have been zero. However, the physical effect of forming/regenerating bound quarkonium is still there, and the same statement can be made for dissociation. Therefore, in order to be able to interpret quarkonium suppression data in terms of an underlying quantum field theory that can be strongly coupled, it will be necessary to have a transport formalism that can account for both Markovian and non-Markovian effects. This is a brave new challenge calling for new theory developments, with the potential of informing us further about the microscopic structure of the QGP.

This work is supported by the U.S.~Department of Energy, Office of Science, Office of Nuclear Physics under grant Contract Number DE-SC0011090\@. X.Y. is supported by the U.S. Department of Energy, Office of Science, Office of Nuclear Physics, InQubator for Quantum Simulation (IQuS) (https://iqus.uw.edu) under Award Number DOE (NP) Award DE-SC0020970 via the program on Quantum Horizons: QIS Research and Innovation for Nuclear Science.

\end{document}